# Pbar Beam Stacking in the Recycler by Longitudinal Phase-space Coating

C. M. Bhat

Fermi National Accelerator Laboratory, P.O.Box 500, Batavia, IL 60510, USA

Phone : 630-840-4821
Fax : 630-840-8461
Email : cbhat@fnal.gov

**Abstract**

Barrier rf buckets have brought about new challenges in longitudinal beam dynamics of charged particle beams in synchrotrons and at the same time led to many new remarkable prospects in beam handling. In this paper, I describe a novel beam stacking scheme for synchrotrons using barrier buckets without any emittance dilution to the beam. First I discuss the general principle of the method, called longitudinal phase-space coating. Multi-particle beam dynamics simulations of the scheme applied to the Recycler, convincingly validates the concepts and feasibility of the method. Then I demonstrate the technique experimentally in the Recycler and also use it in operation. A spin-off of this scheme is its usefulness in mapping the incoherent synchrotron tune spectrum of the beam particles in barrier buckets and producing a clean hollow beam in longitudinal phase space. Both of which are described here in detail with illustrations. The beam stacking scheme presented here is the first of its kind.

# 1. Introduction

Stacking high intensity proton and ion beams in synchrotrons at the same time preserving its emittance, has been one of the major problems for the past several decades. Considerable research has been undertaken at many accelerator laboratories to develop novel stacking schemes [1], viz. *box-car* stacking [2], slip stacking [3], momentum stacking [4], stacking using double harmonic rf systems [5] and transverse and longitudinal phase-space painting [6]. The first four of these use resonant rf systems and the last technique is used while stacking beam from a linear accelerator. Each one of them has its merits and limitations. In a synchrotron like the Fermilab Recycler Ring [7], which exclusively used *barrier rf* systems [8,9] in all of its beam manipulations during Tevatron collider operation, none of the above beam stacking methods could have been used without major rf modifications.

The use of barrier rf in synchrotrons is relatively new to accelerator technology. Significant theoretical as well as experimental research took place during the past two decades particularly due to its use in the Recycler Ring [9-14], induction accelerator at KEK [15], R&D effort at CERN and BNL [16] and the foreseen NESR facility at GSI [17]. The Recycler Ring at Fermilab is an 8 GeV proton/antiproton permanent magnet storage ring. This was used as the primary antiproton depository for beam injection to the Tevatron. The antiproton beam intensity in the Recycler was gradually increased by multiple transfers from the Fermilab Accumulator Ring. Each beam transfer from injection till it is added to the already-existing stack involved a number of rf manipulations. In between transfers, the Recycler beam was cooled using stochastic cooling [18] and electron cooling [19]. It was imperative to keep the emittance of the *cold beam* intact during rf manipulations of beam stacking. Over the past several years, a number of improvements have been made in antiproton stacking schemes in the Recycler [12, 13]. In spite of these, a longitudinal emittance dilution of 10-15% per transfer was observed. In the case of consecutive two or more beam transfers the overall emittance growth was as high as 50%; majority of which was attributed to the rf manipulations involved in these schemes. As a result of these issues, further improvements in beam stacking were in high demand.

In this paper I present a novel scheme of beam stacking called "longitudinal phase-space coating" (LPSC) [20]. The method of beam coating explained here is different from longitudinal phase space painting previously explained in the literature [6]. The longitudinal phase space density of the initial (cold) beam can be held constant for any number of consecutive beam transfers and the emittance growth for the newly arrived beam will be minimal. The presence of barrier rf buckets in the ring is critical to use this novel technique. We describe the working principle, multi-particle beam dynamics simulation to convincingly validate the principle, an experimental demonstration of the LPSC using nearly rectangular barrier pulses in the Recycler, an application of this method to measure synchrotron spectrum of beam in a barrier bucket and finally creating hollow beam in longitudinal phase space. The beam stacking illustrated here can use any of the barrier rf waveforms illustrated in ref. 10.

## 2. The principle of longitudinal phase-space coating

A barrier rf bucket in a synchrotron is generated either by using a broad band rf system or by a set of fast kickers which produce a minimum of three regions per revolution period $T_0$ namely a positive and negative voltage barriers with a zero kick region in the pulse gap. Longitudinal dynamics of a charged particle in such an rf bucket is characterized by its energy offset $\Delta E$ from synchronous energy $E_0$ and a time coordinate $\tau$. (The time coordinate is selected relative to a fixed phase point in the rf wave; generally relative to the center of the bucket). Such a particle will continue to slip relative to a synchronous particle in the region with zero rf voltage. It will lose or gain energy as soon as it encounters a barrier pulse and this will continue until there is enough kick from the barrier pulse to change its direction of slip. Thus, the barrier buckets sets the particles into synchrotron oscillations. Then the equation of motion of any particle in a synchrotron is given by [10],

$$\frac{d\tau}{dt} = -\eta \frac{2\pi \Delta E}{T_0 \beta^2 E_0} \quad \text{and} \quad \frac{d(\Delta E)}{dt} = \frac{eV(\tau)}{T_0} \tag{1}$$

The quantities $e, \eta$ and $\beta$ are electronic charge, phase slip factor and the ratio of the particle velocity to that of light, respectively. $-\tau$ is the time difference between the arrival of the particle and that of a synchronous particle at the center of the rf bucket.

$V(t)$ is the amplitude of the rf voltage waveform. It is important to note that the fractional change in slipping time $\Delta T = [\eta T_D / \beta^2 E_0]\Delta E$, which implies that the slipping time of an off energy particle is proportional to $\Delta E$. Hence, the particles closer to the synchronous particle have a longer synchrotron oscillation period. From Eqs. (1) we can obtain the general Hamiltonian for synchrotron motion for an arbitrary barrier rf wave form as,

$$H(\Delta E, \tau) = -\frac{\eta}{2\beta^2 E_0}\Delta E^2 - \frac{e}{T_0}\int_0^\tau V(t)dt \tag{2}$$

The second term in the above equation represents the potential energy of the particle. In the absence of *intra-beam scattering* and *synchro-betatron coupling* a particle will continue to follow the contour of a constant Hamiltonian as it oscillates in an rf bucket. It can be shown that the maximum value of the energy offset, $\hat{\Delta E}$, of a particle during its synchrotron motion in a barrier bucket is related to its penetration depth $\hat{T}$ into the barrier by,

$$\hat{\Delta E} = \sqrt{\frac{2\beta^2 E_0}{|\eta|T_0}\left|\int_{T_2/2}^{T_2/2+\hat{T}} eV(t)dt\right|} \tag{3}.$$

$T_2$ is the pulse gap. The $\hat{\Delta E}$ represents the bucket height when $\hat{T}$ the total width of the barrier pulse assuming anti-symmetric barrier pulses with respect to the center of the bucket. For an ideal rectangular barrier bucket one can replace $\int_{T_2/2}^{T_2/2+\hat{T}} eV(t)dt = eV_0 \hat{T}$ which simplifies Eq. (2) and (3) considerably. In this case it is easy to imagine that any barrier bucket can be looked upon as one or more barrier buckets, one inside the other, so that one of the inner one confines all the particles whose maximum energy offset is below $\hat{\Delta E}$ with a clear boundary. (This is because with a barrier rf bucket in a synchrotron the concept of *harmonic number* does not exist.)

The principal goal of the new stacking scheme is to isolate particles of certain maximum energy spread using an inner barrier bucket (*mini-barrier* bucket). The maximum potential energy of these particles is set at the same level (or slightly above) as that of the minimum potential energy of the newly arriving particles.   Thus, one can coat the injected beam on the top of the isolated particles. The coating takes place in $(\Delta E, \tau)-$ space. The particles in the mini-bucket will be left undisturbed throughout the stacking.

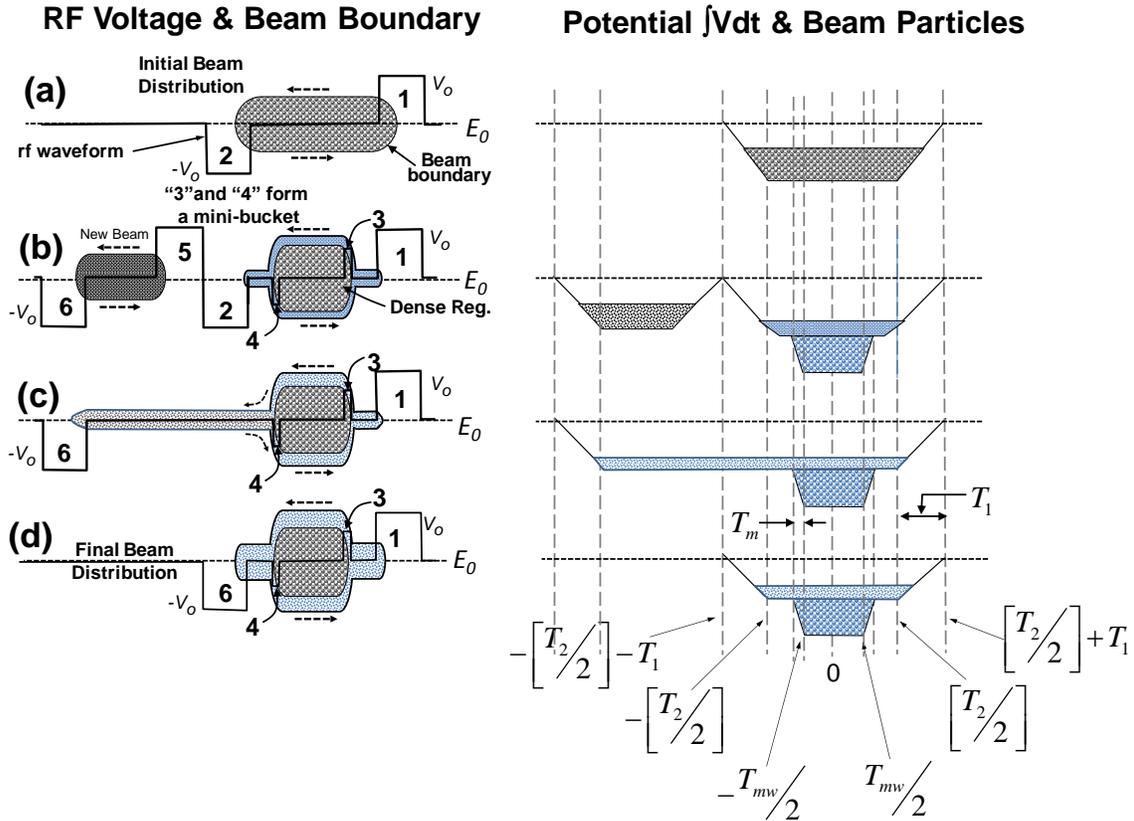

Figure 1: Schematic view of beam stacking by LPSC. The phase space (left) and potential diagrams (right) are shown for different stages of stacking: (a) original beam, (b) after capturing a part of the original beam in a mini-barrier bucket and injection of a new beam, (c) a stage of coating of the new beam on top of the original beam after removal of barrier pulses "2" and "5", and (d) after coating. The voltage wave forms (solid lines) and direction of the synchrotron motion of the beam particles in longitudinal phase-space are also shown in each case (left figures). The horizontal line indicates time axis.

A schematic view of various stages of the LPSC scheme for a synchrotron storage ring with the corresponding rf wave forms, the beam phase space boundaries and the beam particles in the potential well are shown in Fig. 1. Here one assumes that the synchrotron is operating below its transition energy.  The initial beam distribution is shown in Fig. 1(a). Before the transfer of new beam, a mini-bucket made of two barrier pulses "3" and "4" is adiabatically opened as shown in Fig. 1(b). The mini-bucket isolates particles in a phase space area $\varepsilon_m = 2T_{mw}\Delta E_m + 4T_0|\eta|\Delta E_m^3/[3\beta^2 E_0 eV_m]$. The quantities $T_{mw}$, $\Delta E_m$ and $V_m$ are pulse gap, maximum energy spread and pulse height for the mini-bucket respectively. $T_m$ in the Fig. 1 represents the pulse width of the mini-barrier bucket. One can see that if $\varepsilon_m$ is chosen to be total area of the initial beam then the coating takes place on the boundary of the initial beam. In the illustration shown here we chose $\varepsilon_m$ < longitudinal emittance of the initial beam to keep the case more general. Figure 1(b) also shows newly injected beam in a separate barrier bucket made of rf pulses "5" and"6". For simplicity, the parameters of barrier pulses "5" and "6" are chosen similar to those of "1" and "2", respectively. The rest of the LPSC scheme involves a set of two distinct rf gymnastics for every beam transfer. The barrier pulses "2" and "5" are removed to coat the injected beam as shown in Fig.1(c). As these two barrier pulses are slowly minimized simultaneously, the energy spread of the injected beam will decrease initially symmetric to $E_0$. The particles from the new injection continue to slip along the contours of constant Hamiltonian in the injection bucket until they combine with that of "1" and "2". In the absence of barrier pulses "2" and "5", the newly arrived particles follow new contours around the mini-bucket as in Fig. 1(c). Eventually, the rf pulse "6" is moved to the location of "2" adiabatically to complete the coating process as shown in Fig. 1(d).

## 3. Experimental Demonstration of LPSC in the Fermilab Recycler

The LPSC method of beam stacking described above has been tested in the Recycler. The Recycler operates below the transition energy of 20.27 GeV and has $T_0$ = 11.12 μsec. It was equipped with four ferrite loaded broad band barrier rf cavities individually driven by a solid-state power amplifier and capable of providing rf pulses of practically any shape with a maximum amplitude of about 2 kV [9] and a very versatile LLRF control to carry out varieties of rf manipulations [21]. The Tevatron collider demanded maximum of about $400\times10^{10}$ antiprotons to be cooled to less than about 70 eV s for its optimal operation. The barrier bucket parameters were optimized to give total bucket area in excess of 250 eVs (with bucket-area to bunch-area ratio of >2) with enough safety margins for beam stacking and beam extraction [11-13].

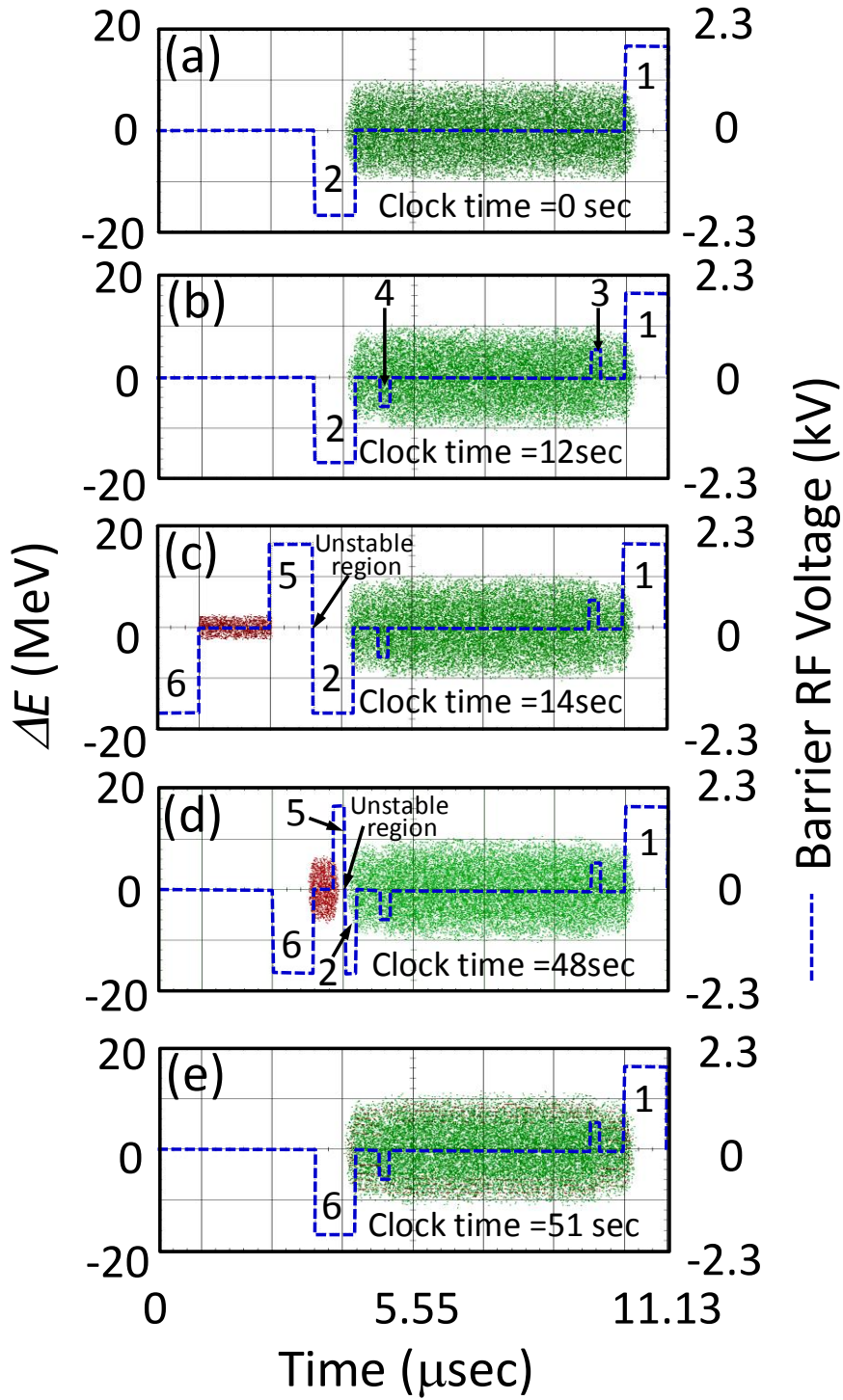

Figure 2: Simulated barrier rf wave form (blue dashed curve) and the phase-space distributions of beam particles in the Recycler for (a) initial, (b) after opening a mini-barrier bucket, c) injection of new beam, d) intermediate stage of coating and, e) after coating. Total clock time to perform the rf manipulations are also shown.

We tested the LPAC scheme in the Recycler in two steps. First, computer simulations using a multi-particle beam dynamics code, ESME [22], were carried out to establish the sequences of rf manipulation. Then, experiments were done with proton beams. Finally, the scheme was implemented operationally in the Recycler to integrate it with the rest of the collider program.

*3.1. Beam dynamics simulations*

Figure 2 shows simulated beam particle distributions in longitudinal phase space along with the barrier rf pulses for the LPSC scheme. These simulations are in the lines of thought of schematic shown in Fig. 1. The initial beam particle distribution was confined in a barrier bucket with a pulse width and height of about 0.91 μsec and 1.93 kV, respectively, and the gap between the two rf pulses was about 5.89 μsec as shown in Fig. 2(a). We chose the total phase space area of the rf bucket to be 250 eVs and that of the initial beam to be about 101 eV s (95%) for illustration. Before populating the new beam, a part of the initial beam was isolated in a mini-barrier bucket (see Fig. 2(b)) of total area equal to 45 eVs inside the initial barrier bucket by opening it iso-adiabatically in about six synchrotron period of the outer most particles of the mini-bucket. The pulse height, width and gap of the mini-bucket were about 0.64 kV, 0.19 μsec and 4.4 μsec, respectively. There are an infinite number of ways to select the rf parameter of the mini-bucket to confine 45 eV s phase space area (the phase space area is a function of pulse height, pulse width and pulse gap, which can be varied in such a way that total area is 45 eVs but pulse height <2 kV, pulse gap <4.8 μsec and pulse width <2.4 μsec). Figure 2(c) shows population of new beam of about 8 eV s in a separate the barrier bucket. The barrier pulse properties for the new beam injection were chosen to be similar to those used for the initial distribution. This makes the rest of the rf manipulations somewhat simpler. Subsequently, the barrier pulses separating the initial distribution and the newly arrived beam distribution were removed by applying morphing technique [12, 13]; the widths of the two rf pulses were reduced simultaneously and symmetrically about the unstable region as indicated in Fig. 2(c). Also, the injected beam was moved towards the initial beam to match energy spread to two beams without changing the distribution of the initial beam. Figure 2(d) shows an intermediate stage of morphing.  We found that this sort of morph merging gave minimal emittance growth to the new beam and took less time.  As the barriers separating the boundary between the new and initial beams become small along with $\int_0^\tau V(t)dt$, the contours of constant Hamiltonian start merging from both side. At the same time the particles from both sides get mixed up and complete the coating. The final distribution is shown in Fig. 2(e). The simulation showed that the final emittance of the beam was about 109 eVs (95%); thus a negligible emittance growth was observed. We have investigated a few variations of the morphing technique presented here. The technique explained above found to give optimum performance. The steps in Figs. 2(c) to 2(e) can be repeated as many times as needed for multiple layers of beam coating.

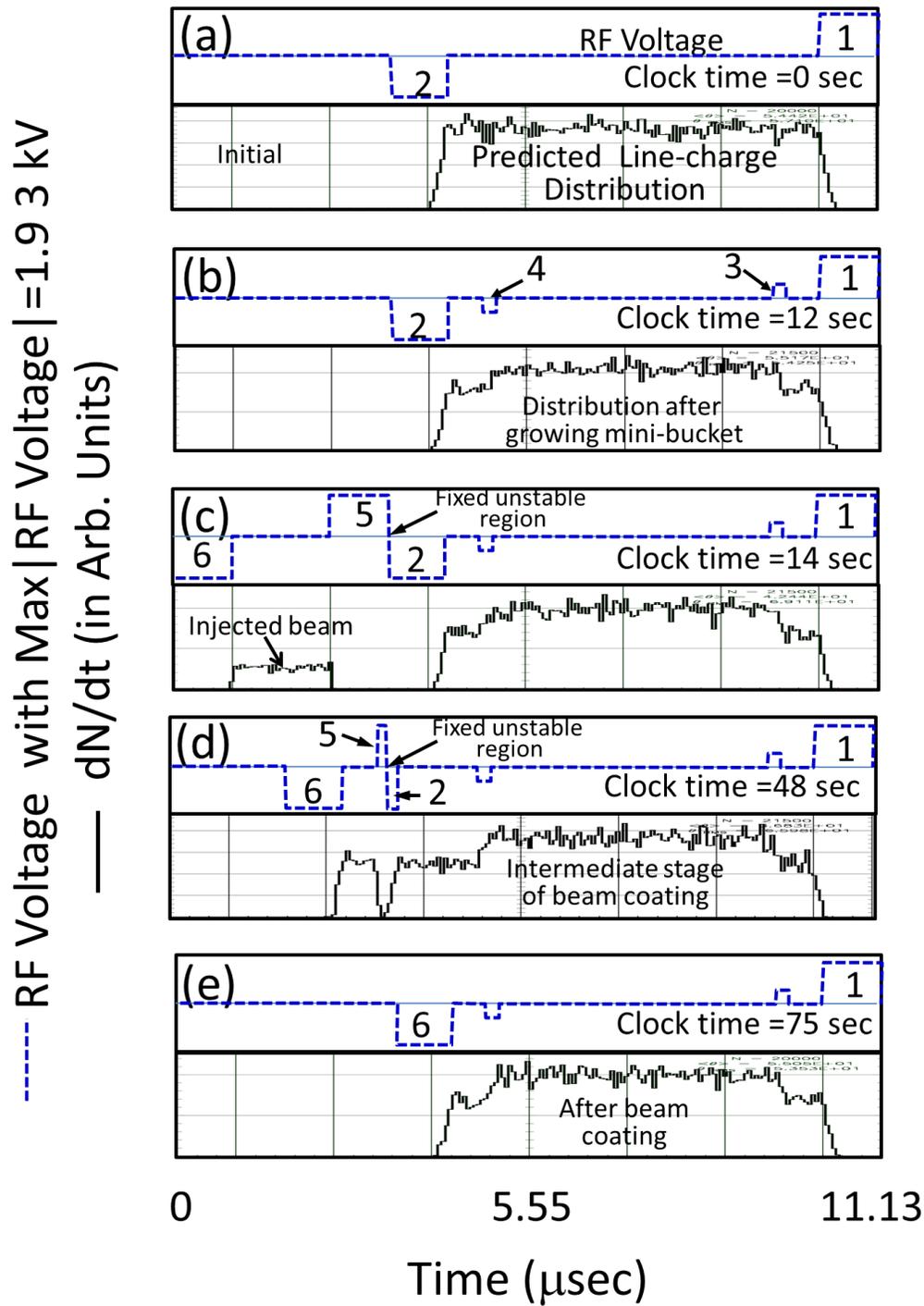

Figure 3: Simulated barrier rf wave form (blue dashed curve) and the line-charge distributions of protons in the Recycler obtained from the time projections of the phase space distributions shown in Fig. 2. For details see the text. The predictions from "b" to "e" are similar to those presented in Figs. 4(A)-a to -d.

Figure 3 depicts the simulated line-charge distributions along with the rf wave form for various stages of coating shown in Fig. 2. Note that there lies some subtle differences between these two cases during the later stages of rf maneuvering. In the case shown in Fig. 2 we observe i) the unstable point moves as the barrier pulse widths are decreased, ii) at the same time the left most barrier pulse of 0.91 μsec cogged towards right side to compress the injected beam in such a way that the length of the initial beam does not change almost until barrier pulses "2" and "5" disappear and furthermore by the time "2" and "5" disappear, the barrier pulse "6" will have moved to the location of "2" in Fig. 2(a)   iii) hence, the coating is done rather fast still adiabatic enough.  In the latter case, i) the unstable point remains fixed till the barrier pulses around it  ("2" and "5") disappear even though the injected beam is being compressed   ii) an additional compression is needed to complete the coating. As a result of this the case shown in Fig. 3 takes about 20 sec longer than that shown in Fig. 2. In any case, the beam experiment was carried out only for the latter case. For example, the line charge distribution in Fig. 3 should be compared with the wall current monitor data measured during the beam experiment shown in Fig. 4(A).

The simulation clearly shows that emittance preservation in this scheme depends on the iso-adiabaticity of the rf manipulation steps just like any other rf gymnastics. The beam particles in the barrier buckets are predicted to have considerable amount of the synchrotron tune spread [10]. As a result of this one has to give special attention while selecting the rate of rf maneuvering.

*3.2. Experimental Demonstration*

The beam test was carried out in the Recycler using proton as well as antiproton beams [12, 20] with varieties of initial beam intensities, different ways of opening the mini-buckets and coating rf manipulations. The other considerations were the length of the mini-bucket a) considerably less than $T_2$ and b) same as the pulse gap of the initial bucket, c) capturing all and part of the initial beam in the mini-bucket. Here I illustrate two cases corresponding to "a" and "b". In both illustrations only a part of the initial beam was captured by the mini-bucket. The scope pictures of the measured wall current monitor data and the corresponding rf wave form at various stages of the beam manipulations are shown in Figs. 4(A) and 4(C).

In the case of the illustration shown in Fig. 4(A) (same as 4(B)), about $3.09 \times 10^{12}$ antiprotons were stored in a rectangular barrier bucket similar to the one illustrated in our simulation demonstrations. The beam was cooled using stochastic cooling as well as electron cooling to a longitudinal emittance ~ 83±8.0 eV s (95%). Then, a mini-bucket with $V_0$ ~0.67 kV, $T_m$ ~0.19 μsec and $T_2 - 2T_m$ ~4.43 μsec having an area ~46 eV s was opened in the middle of the initial beam. The height of the mini-bucket $\Delta E_m$ was 4.9 MeV. The measured line-charge distribution and the Schottky data for the beam at his stage are shown in Fig. 4(A)-a (and 4(B)-b top picture as indicated) and 4(B)-a, respectively. Figure 4(A)-b shows data after the first transfer of about $10 \times 10^{10}$ antiprotons with a longitudinal emittance of 6.7±0.8 eVs into the already opened matched four 2.5 MHz buckets of the Recycler (see also 4(B)-b) . Finally, the newly arrived antiprotons were coated on the antiprotons in the mini-barrier bucket without disturbing it following the rf manipulation procedure depicted in Fig. 3. Figure 4(c) shows the wall current monitor data corresponding to an intermediate stage of beam coating soon after the 2.5 MHz rf waves and the rectangular barrier pulses separating the mini-bucket and newly arrived beam are removed. Completion of 1st coating is shown in Fig. 4(d) (see also 4(B)-c-top trace as indicated). During this experiment we did a total of two beam coatings. The second coating consisted of $7\times10^{10}$ antiprotons and with a longitudinal emittance of 6.0±1.0 eVs. The Schottky data taken after second coating is shown in Fig. 4(B)-d. The longitudinal emittance of the final beam of intensity $3.26\times10^{12}$ antiprotons was 96.4±10 eVs (measured using Schottky data and standard formula for rectangular barrier rf bucket).

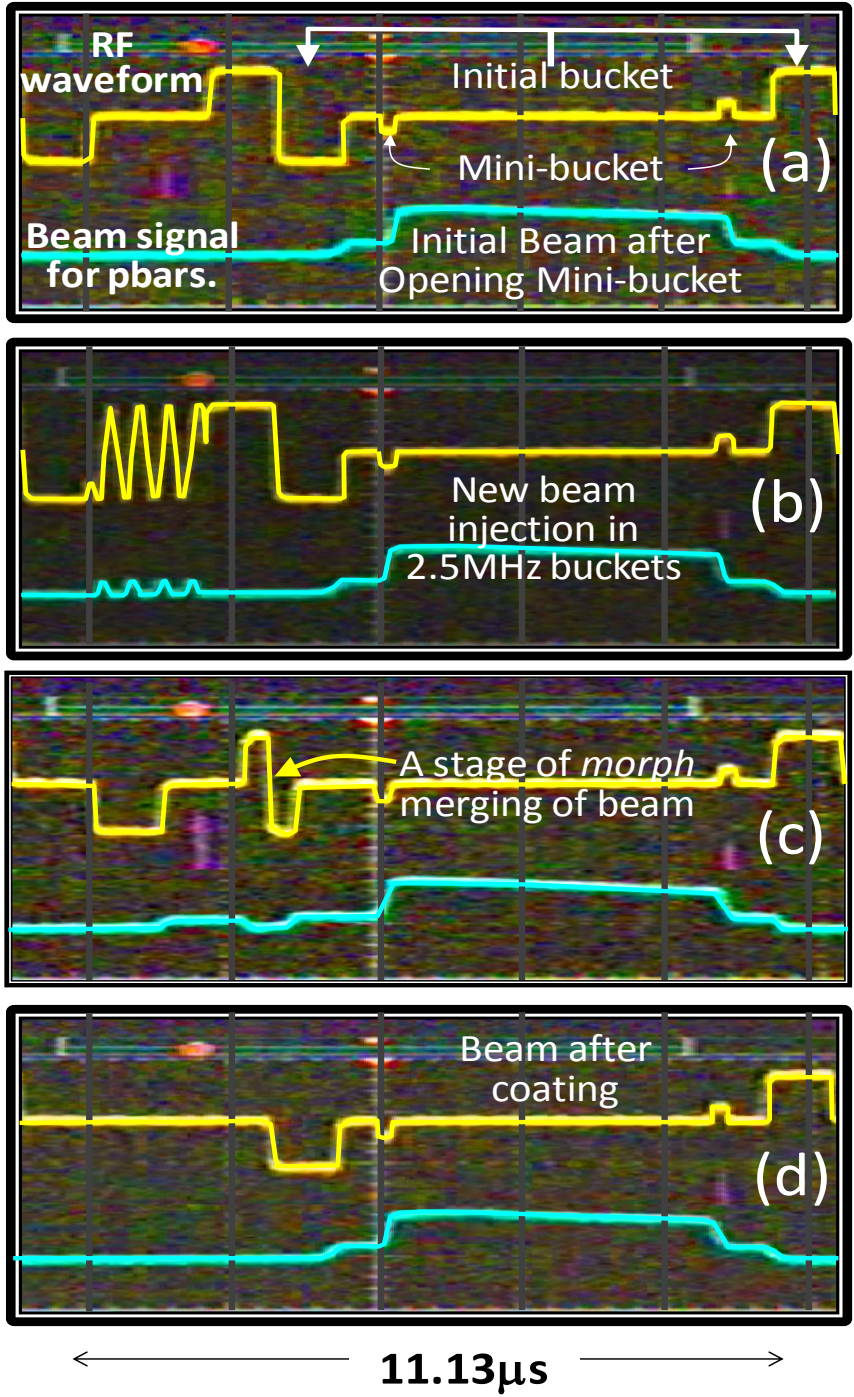

Figure 4(A): The scope pictures for the longitudinal phase space coating in the Recycler with the length of the mini-bucket less than the pulse gap of the initial bucket. The two traces in each of the figures represent rf wave form (top trace) and beam signal from a wall current monitor (bottom trace). The different stages of the coating are shown. Beam intensities in "a" and "d" were about $309\times10^{10}$ and $319\times10^{10}$ antiprotons, respectively. For details see the text.

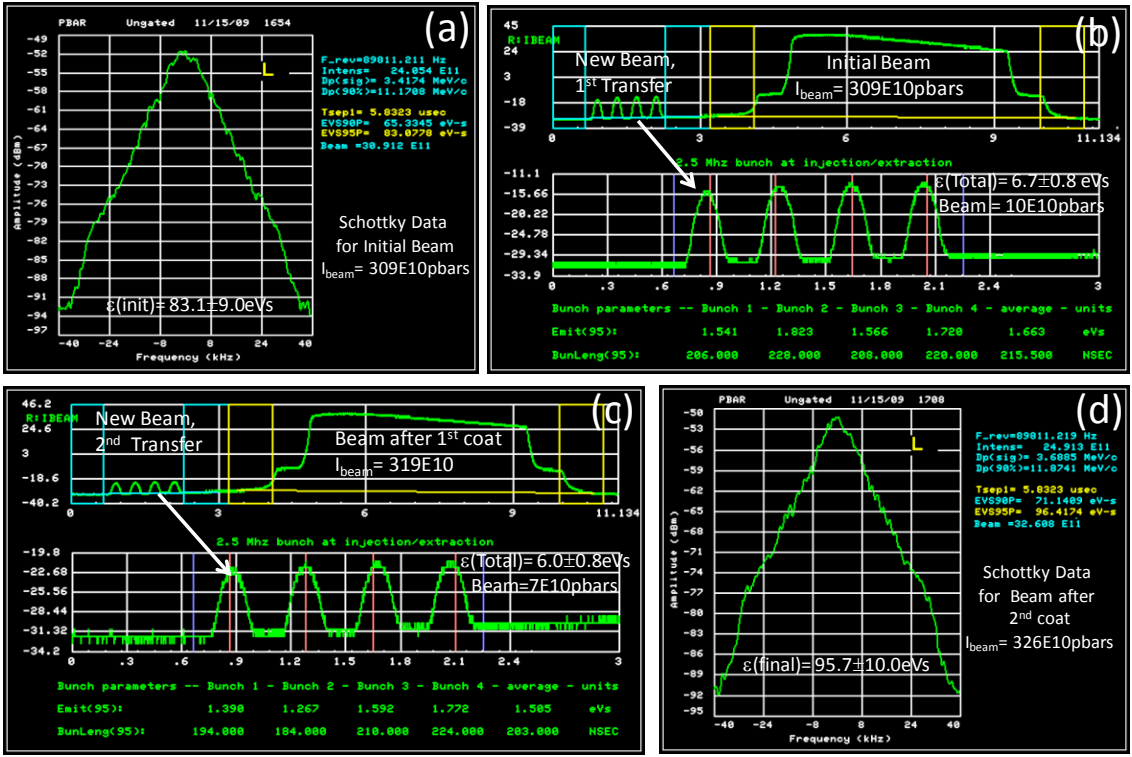

Figure 4(B): a) Schottky data for the initial beam, b) WCM data for initial beam and 1st new injection, c) WCM data for beam after 1st coat (top), and 2nd injection (bottom) and d) Schottky data after 2nd coat. See also Fig. 4(A).

In Figure 4(C) we illustrate an example where the mini-bucket length = $T_2$. In this case the initial beam consisted of about $2.56\times10^{12}$ antiprotons in a barrier bucket similar to one depicted in Fig. 4(A)-a. The beam was cooled to a longitudinal emittance ~

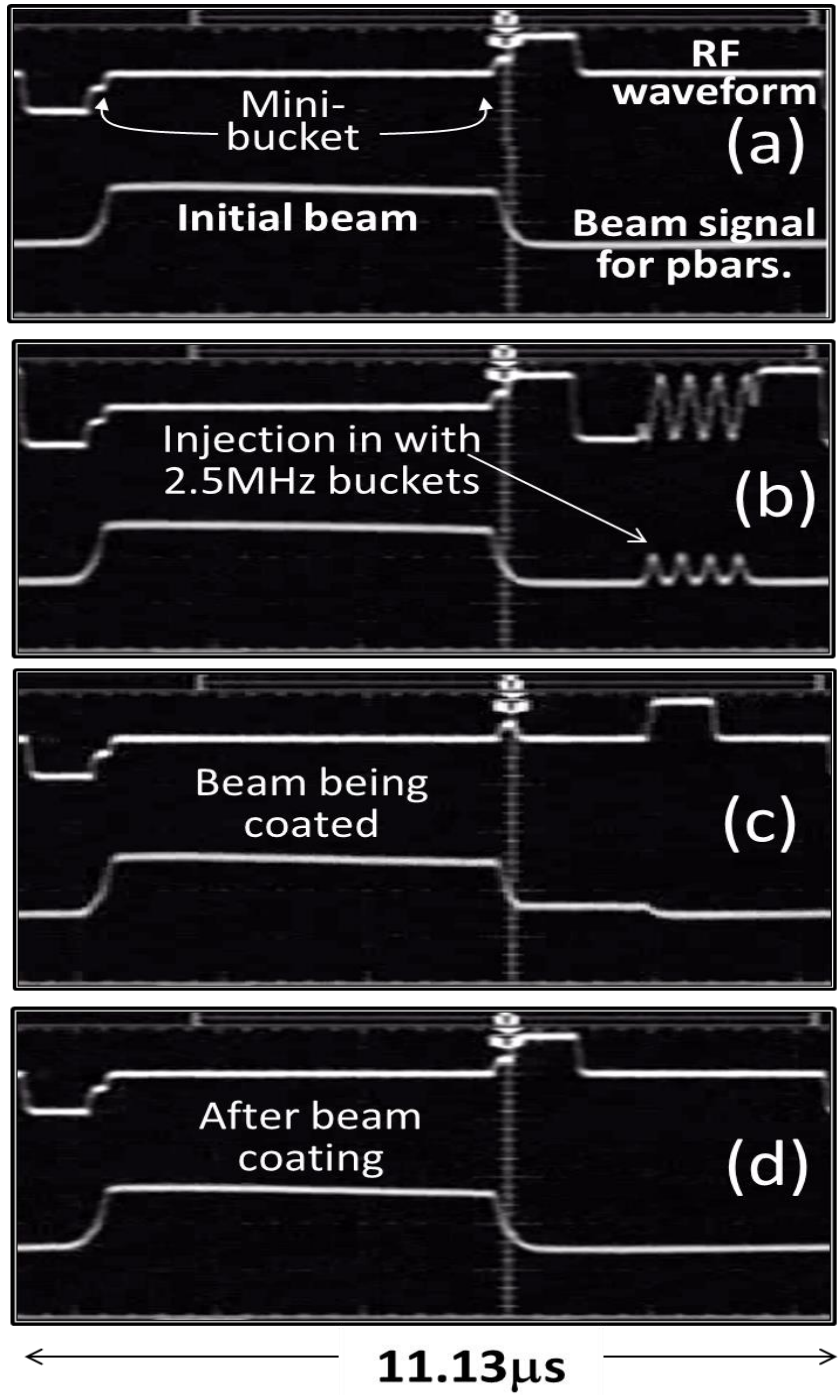

Figure 4**(C)**: The scope pictures for the longitudinal phase space coating in the Recycler. The two traces in each of the figures represent rf wave form (top trace) and beam signal from wall current monitor (bottom trace). The stages of the coating are shown. Beam intensities in "a" and "d" were about $256\times10^{10}$ and $270\times10^{10}$ antiprotons, respectively. For details see the text.

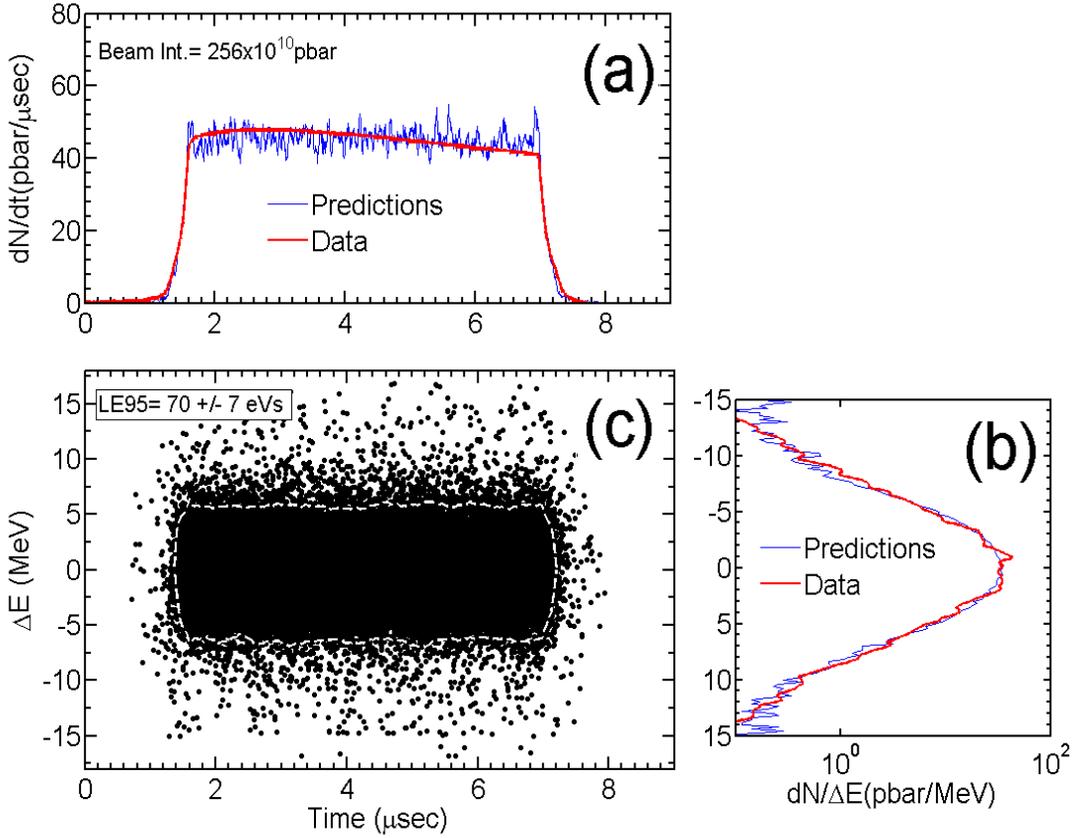

Figure 5: Comparison between ESME simulations with the measurement data for the initial beam: (a) wall current monitor and (b) Schottky detector data. The blue and red traces are, respectively, experimental data and simulations. (c) The simulated longitudinal phase-space distribution of the beam with 95% contour. LE95≈70±7 eVs.

70±7eV s (95%). Then, a mini-bucket with $V_0$ ~0.72 kV, $T_m$~0.25 μsec, $\Delta E_m = 5.8$ MeV and $T_2 - 2T_m$ ~5.4 μsec having an area ~66 eV s was opened to capture about 93% of the initial beam. In this case we had three consecutive coating (separated by ~1 min); the longitudinal emittances and beam intensities for these three coatings were 7±1, 8±1 and 7±1 eV s with $14\times10^{10}$, $9\times10^{10}$ and $5\times10^{10}$ antiprotons, respectively. Figure 4(C)-d shows scope data after the completion of 1st coating. The measured wall current monitor and the Schottky data for the initial beam and after the three coats are shown in Figs 5 and 6 respectively (red traces). The data shown in Fig. 6 corresponds to a total beam of $2.84\times10^{12}$ antiprotons.

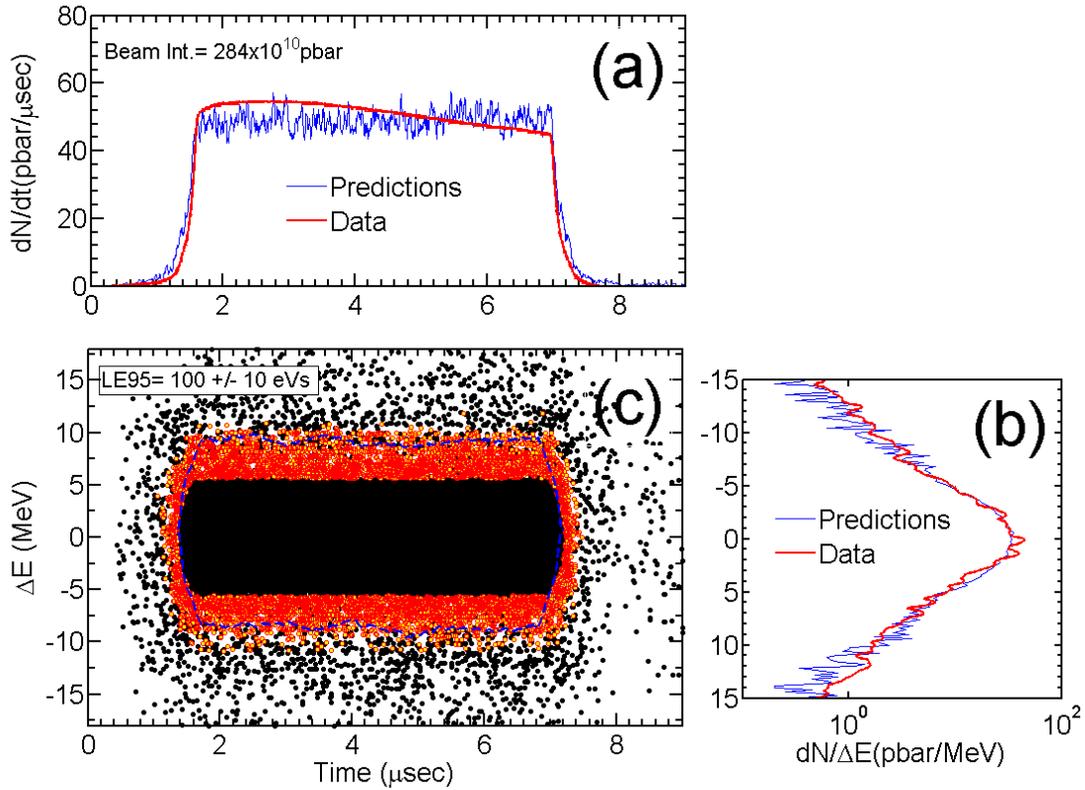

Figure 6: Comparison between ESME simulations and the measurement data for the pbar beam after three coats. The descriptions for three plots and the traces are the similar to that in Figure 5. The beam captured using the mini-barrier bucket can be seen clearly in the middle. The particles outside the mini-bucket mixed well with the newly arrived particles.

Generally, measurement of longitudinal emittance of a beam in a barrier bucket is not straightforward for barrier pulses deviating from a standard rectangular shape. The measurements in the Recycler showed that fan-back signals of a rectangular barrier pulses deviate noticeably from ideal shapes. Further, the combined shape of two rectangular barrier pulses of different heights used in coating, were certainly not rectangular in shape. For example the barrier rf pulses shown in Fig. 4(C) looked like step functions. In addition to this, a) use of a finite number of Fourier components to create a rectangular shape led to non-symmetric rising and falling edges to these barriers with a rise and fall time in the range of 5 to 15 nsec and, b) there was a polar asymmetry for the barrier pulses; the negative pulses were a few percent smaller than the positive barrier pulses (see for example Fig 8(a)). Consequently, the use of standard analytical formula for rectangular barrier bucket given in ref. 10 may not be adequate. Hence, we used a beam Monte Carlo (MC) method [23] in the determination of longitudinal emittance. One of the important requirements to apply standard analytical formula or MC method to estimate longitudinal emittance is that the beam should have reached equilibrium after completion of rf manipulations, which we believe in these cases. We use the ESME code to construct the beam particle distribution in ($\Delta E, \tau$) –space by matching simulated time and energy projections to the measured wall current monitor and Schottky data, respectively, as shown in Figs. 5 and 6 (red and blue traces represent predictions and measurements, respectively). We used measured rf wave form in our MC simulations. The longitudinal emittances for the beam after three coats were found to be 100±10 eV s. The closed contours in Figs. 5(c) and 6(c) represent 95% of the phase-space area of interest. By adding the errors in quadrature, we find that the observed emittance dilution is within measurement errors of about 10% of the experiment.

Figure 7 shows the measurement data in the Recycler during regular collider operation. In this example, we adopted "normal stacking" that involved only the morph merging of the injected beam (which potentially had high risk of longitudinal emittance dilution for the dense cold region of the beam particle because of complete removal of the rf pulses at the time of merging the initial beam and injected beam) up to an intensity of $270 \times 10^{10}$ antiprotons and the rest with LPSC scheme. The vertical dashed line separates these two as indicated in the figure. Each main step in the beam intensity (magenta curve) comprised of a set of three or four beam transfers separated by ~1 min (as indicated in the Fig. 7). The measured average transverse emittance $<\varepsilon_\perp>$, average beam brightness $<d> = N(\times 10^{10})/[\varepsilon_\perp(\mu m)\varepsilon(eVs)]$ and antiproton beam intensity show similar steps as the stacking progressed. Between two successive set of beam transfers the time gap was about forty-five minutes and the beam was cooled mainly using stochastic cooling with about a few minutes of electron cooling only if the beam was not cooled enough to <90 eV s and a transverse emittance of about 2.5 μm. The inset in Fig. 7 is Schottky data measured soon after the final beam coating but with mini-buckets removed. At the end of the beam stacking the antiprotons was used for the collider operation. We also had similar beam stacking with the LPSC scheme at the early part of the beam stacking until the beam intensity reached $> 270 \times 10^{10}$ followed by normal stacking to reach the total intensity of about $400 \times 10^{10}$. In both cases the LPSC scheme was transparent to the rest of the collider operation. A comparison of the final distributions between normal stacking and that obtained from the LPSC showed similar behavior within the measurement errors. The central dense region of the beam distribution was disturbed very little even in the case of normal stacking because the average synchrotron oscillation period for the beam particles close to synchronous energy were in the range of several seconds whereas the rf manipulation was relatively fast. As a result of this, we did not find much difference between these two techniques in the Recycler when the initial beam was extremely cold.

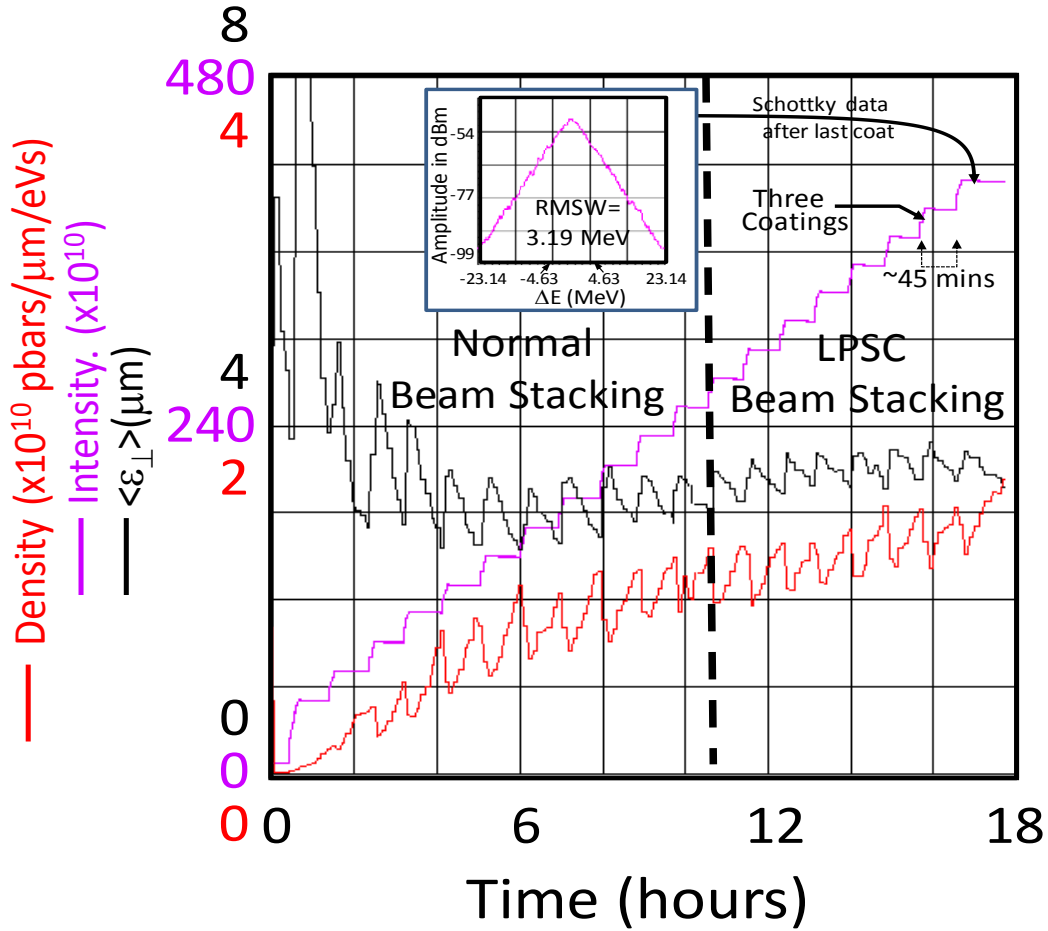

Figure 7: Antiproton stacking before and after implementation of LPSC scheme in the Recycler. The dashed line represents transition from standard beam stacking and the LPSC. The inset is Schottky data measured at the end of beam stacking after mini-bucket was removed.

**4. LPSC to measure incoherent spectrum of beam in a barrier bucket**

One of the essential steps in quantitative understanding of single particle dynamics in an rf bucket involves measurement of synchrotron tune. In the past, such a measurement was made for particles in a sinusoidal rf bucket at the IUCF cooler [24]. The LPSC method of beam stacking provides an elegant method to measure the synchrotron tune spectrum of beam particles in barrier buckets. The incoherent synchrotron frequency $f_s$ of beam particles on the outermost separatrix of mini-rectangular barrier bucket is $f_s^{-1} = 2(T_2 - 2T_m)\beta^2 E_0 /|\eta \Delta E_m| + 4T_0 |\Delta E_m|/eV_m$. For the purpose of illustration a rectangular barrier bucket of pulse height $V_0$=1.84 kV, pulse

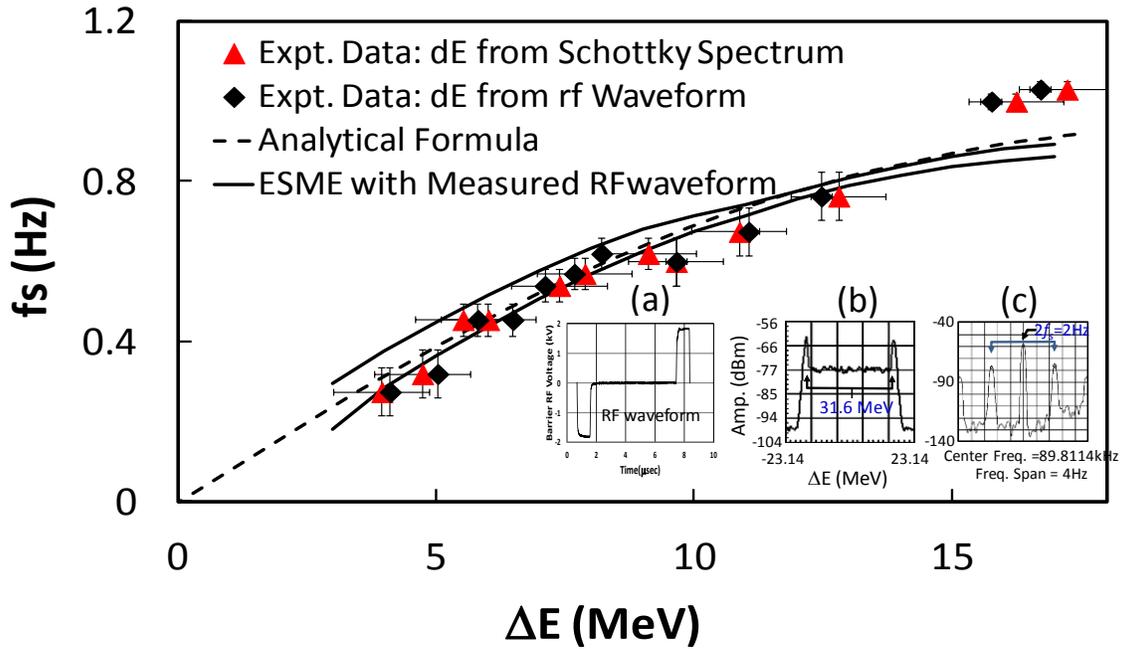

Figure 8: Measured and calculated synchrotron frequency as a function of $\hat{\Delta E}$ of beam particles in a barrier bucket with $V_0$=1.84 kV, $\hat{T}$=0.9 µs and $T_2$=5.9 µs. The dashed line is obtained with analytical formula assuming ractangular barrier waveform. The insets (a) shows the exact rf waveform and, (b) and (c) show Schottky and VCA data for the 15.8 MeV data point, respectively as an illustration.

width $\hat{T}$ =0.91 μs and pulse gap $T_2$=5.9 μsec was chosen as the bucket of interest to map the synchrotron frequency spectrum (see Fig. 8(a)). With no beam in the Recycler, a mini-barrier bucket that occupies the entire pulse gap $T_2$ was grown inside the main bucket (as illustrated in Fig. 1(b)). Then a small amount of proton beam (~$15\times10^{10}$) was coated on to the empty mini-bucket. This created a long bunch with an ideal hole in the central region of longitudinal phase space. The separatrix of the mini-bucket acts as the boundary between the empty region and the coating. The wall current monitor signal was fed to Agilent 89441A 2.65GHz VSA (with a frequency span of 0-4Hz, centered at the Recycler Ring revolution frequency) to measure the synchrotron frequency of the particles sitting on the separatrix. An illustration of measurement data corresponding to $\Delta E_m$=15.8 MeV and $f_s$=1 Hz are shown in insets Fig. 8(b) and Fig. 8(c). By changing the parameters of the mini-bucket and new coatings the entire synchrotron spectrum was scanned. The $\Delta E_m$ in each case was measured using Schottky measurement. The measured synchrotron frequency as a function of $\Delta E_m$ is shown in Fig. 8 along with analytical predictions assuming rectangular barrier pulses and ESME simulations with measured barrier rf wave from Fig. 8(a). We have also shown the $\Delta E_m$ calculated using measured barrier rf wave form and Eq. 3 for comparison. The agreement between them is rather good. The level of discrepancy between measurements and the predictions of synchrotron frequency spectrum can be understood as being due to the shapes of rf pulses used in the experiment and systematic errors in the measured rf voltage versus that used in the calculations.

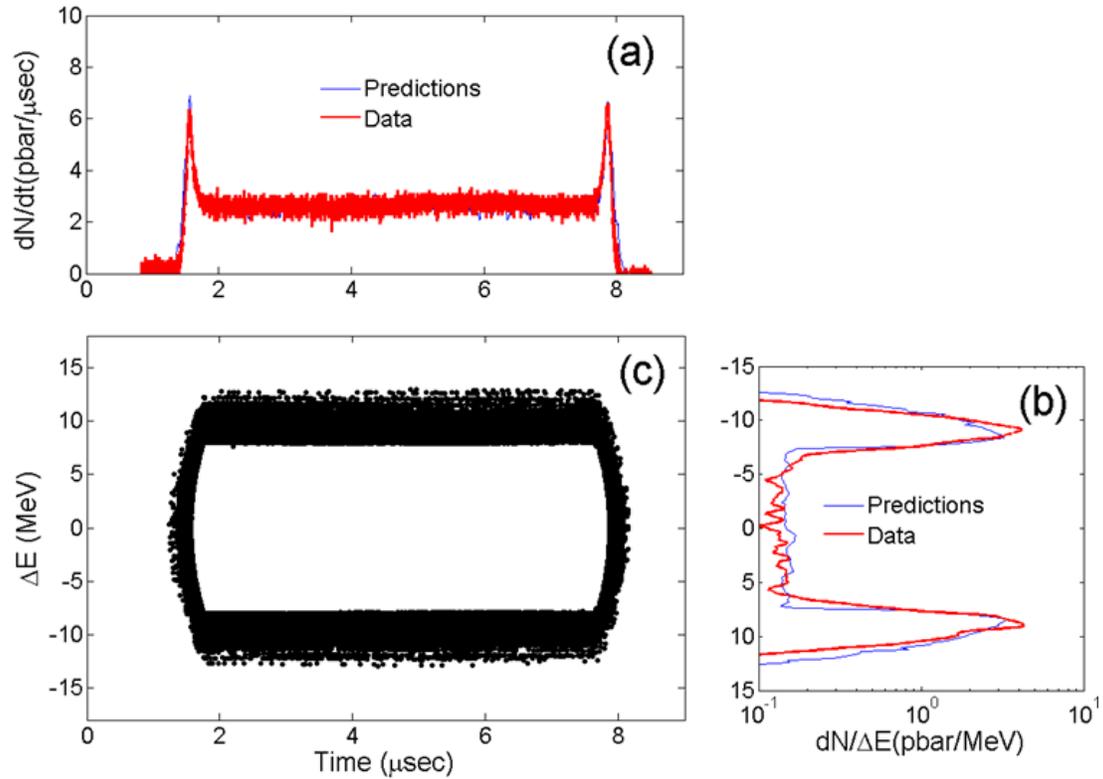

Figure 9: Measurement on a hollow beam. The descriptions of the figures are similar to that presented in Fig.5.

During each of the measurements mentioned above a hollow bunch is created. However, the hole in the longitudinal phase space is maintained by means of mini-bucket. We found that even in the absence of mini-barrier bucket a clean hollow beam can be maintained as shown in Fig. 9. We certainly observe some leakage of beam particles into the hollow region mainly because of non-adiabaticity of the rf manipulation while removing the mini-bucket. Very little degradation in hollow beam is seen even after a long time (of the order of hours). Figures 9 (a) and (b) show measured and ESME predicted line-charge distributions for the hollow beam. The corresponding reconstructed beam particle distribution in the longitudinal phase-space is shown in Fig. 9(c). We see quite good agreement in the predictions and the measurement data.

## 5. High intensity effect

The simulations presented in Figs. 2 and 3 are using single-particle beam dynamics without including beam space-charge or wake fields effects. Generally, these simulations are sufficient to illustrate the proof of principle. However, collective effects in simulations are quite important in understanding the high intensity behavior. The results presented in Figs. 5 and 6 which estimate the longitudinal emittance for the initial and final distributions include multi-particle effects, namely, 1) space charge, 2) reaction of the beam environment (*e. g.,* beam pipe) on the beam distribution and 3) rf cavity impedance. The total impedance $Z(\omega)$ seen by a Fourier component of the beam current at frequency $\omega/2\pi$ is modeled as

$$\frac{Z(\omega)}{n} = -j\frac{Z_0}{2\beta\gamma^2}\left[1 + 2\ln\frac{b}{a}\right] + \frac{Z_{\parallel}(\omega)}{n} + \frac{Z_{Ind}}{n} \tag{4}$$

in ESME. $a$ and $b$ are average beam size and beam pipe size, respectively. The first term in Eq.4 represents the longitudinal space charge impedance with $Z_0 = 377\,\Omega$ (impedance of free space). $Z_{\parallel}$ is the longitudinal coupling impedance of resonant structures, approximated by simple resonance, $Z_{\parallel}(\omega) = R_{shunt}/\{1 + jQ[(\omega_r/\omega) - (\omega/\omega_r)]\}$ with $\omega_r = 2\pi/T_0$. The shunt impedance $R_{shunt}$, for all four cavities together was about 200 $\Omega$ and $Q = 1$ [9]. $Z_{Ind}$ is the total wall impedance of the beam pipe. For long bunches the impedance is mainly inductive. We model $Z_{Ind}(\omega)/n = j(Z/n)(\omega/\omega_r)$ with $Z/n = 0.1\,\Omega$ for the entire ring and the frequency $\omega$ is changed in the range 0-4.4 GHz. The simulations showed that the second term in Eq. 4 plays a very important role, and gave rise to significant asymmetry in a long bunch because of potential well distortion. Experimentally, this phenomenon was observed in the Recycler on long bunches at beam intensity as low as $20\times10^{10}$ protons [25]. This posed a serious problem for the collider operation which demanded flat long bunch in the Recycler which can provide equal intensity antiproton bunches after longitudinal momentum mining [11]. Hence, to address this problem once and for all, a FPGA-based adaptive correction system was implemented in the Recycler LLRF [26] and was tested to intensities in excess of $500\times10^{10}$. Consequently, in our simulations mentioned above we assume full compensation. Further simulations showed that the LPSC scheme can be used in the Recycler without any detrimental effects even beyond $600\times10^{10}$ antiprotons, which is about 2.5 times the original design intensity [7].

## 5. Summary

We have proposed and validated a novel beam stacking method, longitudinal phase space coating, for a storage ring that uses rf barrier buckets. The scheme has been studied using multi-particle beam dynamics simulations and we have illustrated the technique with beam experiments in the Recycler. This method was also been successfully implemented and tested in the Recycler during the Fermilab collider operation. The method works in such a way that the majority of the central region of the phase space is undisturbed throughout the stacking. We have demonstrated that this technique can give less than 10% emittance dilution during antiproton stacking.

A spinoff of the LPSC scheme is its use in measuring the incoherent synchrotron spectrum of the beam distribution in a barrier bucket. We illustrated such a measurement on one of the barrier buckets used in the Recycler. The measurement data is reproduced quite well by an analytical calculation and ESME simulations. The LPSC technique is used to create an ideal hollow beam in longitudinal phase space. At this time, such a hollow beam is purely of academic interest. In the future, this may be of very high interest in the context of studying varieties of distribution functions, and to beam physics generally.

As a final note, we expect that the applications of the technique described here may not be unique to high-energy storage rings for protons and antiprotons, but may be very well useful in heavy ion storage rings. We believe it should also have broad applications in other low energy circular storage rings as well.

## Acknowledgments

The author would like to thank D. Neuffer, D. Wildman and Jim MacLachlan for many early discussions. Special thanks are due to Shreyas Bhat for careful editing of the manuscript. Cans Gattuso and the Fermilab Accelerator Division MCR crew are also acknowledged for their help during the beam experiments. This work is supported by Fermi Research Alliance, LLC under Contract No. DE-AC02-07CH11359 with the U.S. Department of Energy.